\documentclass{article}

\usepackage[total={6in, 8in}]{geometry}
\usepackage{authblk}
\usepackage{graphicx}
\usepackage{amssymb}
\usepackage{amsmath}
\usepackage{color}
\usepackage{url}
\usepackage{hyperref}
\usepackage{subcaption}
\usepackage{float}

\title{Relevant, Hidden, and Frustrated Information in High-Dimensional Analyses of Complex Dynamical Systems with Internal Noise} 

\author[1]{Chiara Lionello}
\author[1]{Matteo Becchi}
\author[1]{Simone Martino}
\author[1]{Giovanni M. Pavan\thanks{Corresponding author: giovanni.pavan@polito.it}}

\affil[1]{Department of Applied Science and Technology, Politecnico di Torino, Torino 10129, Italy}

\date{\today}
\usepackage[style=numeric, maxnames=1, minnames=1, sorting=none]{biblatex} 
\begin{document}

\maketitle

\begin{abstract} 
Extracting from trajectory data meaningful information to understand complex molecular systems might be non-trivial. High-dimensional analyses are typically assumed to be desirable, if not required, to prevent losing important information. But to what extent such high-dimensionality is really needed/beneficial often remains unclear. Here we challenge such a fundamental general problem. As a representative case of a system with internal dynamical complexity, we study atomistic molecular dynamics trajectories of liquid water and ice coexisting in dynamical equilibrium at the solid/liquid transition temperature. To attain an intrinsically high-dimensional analysis, we use as an example the Smooth Overlap of Atomic Positions (SOAP) descriptor, obtaining a large dataset containing $2.56\times10^6$ 576-dimensional SOAP vectors that we analyze in various ways. Our results demonstrate how the time-series data contained in one single SOAP dimension accounting only $<0.001\%$ of the total dataset's variance (neglected and discarded in typical variance-based dimensionality-reduction approaches) allows resolving a remarkable amount of information, classifying/discriminating the bulk of water and ice phases, as well as two solid-interface and liquid-interface layers as four statistically distinct dynamical molecular environments. Adding more dimensions to this one is found not only ineffective but even detrimental to the analysis due to recurrent negligible-information/non-negligible-noise additions and ``frustrated information" phenomena leading to information loss. Such effects are proven general and are observed also in completely different systems and descriptors’ combinations. This shows how high-dimensional analyses are not necessarily better than low-dimensional ones to elucidate the internal complexity of physical/chemical systems, especially when these are characterized by non-negligible internal noise.

\end{abstract}

\newpage

\section*{Introduction}

Elucidating the physics of complex dynamical systems is a challenging task and remains one of the most thought-provoking topics in both physics and chemical physics. These systems feature multiple levels of local and collective dynamical events that coexist and interconnect within complex dynamical networks, making their understanding and classification particularly difficult. Phenomena ranging from microscopic events -- such as phase transitions, nucleation processes~\cite{wolde1997enhancement, lutsko2019crystals}, or the intricate internal dynamics of molecular assemblies~\cite{hagan2021equilibrium, crippa2022molecular} -- to macroscopic events like the rapid turns and convolutions of large crowds of individuals, bird flocks~\cite{cavagna2010scale, nagy2010hierarchical}, and fish schools~\cite{porfiri2018inferring}, highlight the need for advanced analytical methods to detect and track the diverse dynamical environments that emerge within these systems. 

A common assumption when studying complex systems (of any scale), about which little is known \textit{a priori}, is that high-dimensional analyses are desirable, and even necessary, to avoid losing information due to prior assumptions or incomplete characterization. However, selecting a set of physically relevant descriptors capable of providing a complete characterization of a specific system requires a deep knowledge of the system itself, which is not always available. Moreover, managing high-dimensional analyses and extracting meaningful physical insights from them is often challenging. Fundamental questions arise: What is the intrinsic dimension of the dataset~\cite{facco2017estimating, dinoia2024beyond}? Which components are the most relevant? 

At the molecular scale, high-dimensional analyses can be approached in various ways. One common method is to use descriptors that transform, for example, the simulation trajectories of the system’s units into analyzable data formats, such as time-series. Multiple descriptors can be employed under the assumption that each captures different (orthogonal) information. However, the degree to which the information provided by these descriptors is complementary or redundant is not known \textit{a priori}. To ensure information-rich analyses, descriptors that are intrinsically high-dimensional by their mathematical definition are particularly valuable. A notable example is the Smooth Overlap of Atomic Position (SOAP) descriptor~ \cite{bartok2013representing}, which, inspired by wave functions and orbital definitions in quantum mechanics, effectively identifies local structural environments in complex molecular systems. 

Similarly to other descriptors, such as atomic cluster expansions (ACE)~\cite{drautz2019atomic} or N-body iterative contraction of equivariants (NICE)~\cite{nigam_recursive_2020}, SOAP provides high-dimensional mathematical representations of local density and order/disorder. Recently, SOAP has been successfully applied to investigate complex molecular systems, including liquid-solid and liquid-liquid coexistence in aqueous systems~\cite{Donkor2024spatavg, offeidanso2022, Ansari2020, Martino2024}, and non-trivial dynamics and local disorder events in metal surfaces and nanoparticles~\cite{Caruso2023tsoap, Caruso2024, crippa2023machine, Rapetti2023, Cioni2024,perrone2024}, and ionic behaviors in dense environments~\cite{Lionello2022}. 

In principle, the high-dimensional description provided by SOAP analyses can be extremely useful, especially for studying systems about which little is known \textit{a priori}~\cite{bartok2013representing,Lin2024}. However, while high-dimensional descriptions are often assumed to be desirable for such systems, our recent work has shown that in many real cases -- including aqueous, solid, and metallic systems of various kinds~\cite{Caruso2023tsoap} -- a mono-dimensional version of SOAP, called \textit{Time}SOAP~\cite{Caruso2023tsoap},  can extract more meaningful features than the standard high-dimensional SOAP analysis by tracking the variation of the SOAP spectrum of each molecule over time. 

This finding raises several important questions: To what extent is high-dimensionality truly necessary to capture the physics of these systems? How can a high-dimensional analysis be less informative than its dimensionally-reduced counterparts? What type of information remains hidden, where is it located, and how can it be extracted? Is the information contained in the orthogonal dimensions of a high-dimensional feature space always additive, does adding a second dimension always contribute new information? 
These questions are particularly central when using intrinsically high-dimensional descriptors such as, \textit{e.g.}, SOAP or ACE~\cite{drautz2019atomic}, but they also apply to any combination of descriptors, whether physics-inspired, abstract, or general. At their core, these questions challenge a fundamental issue in physics: How to effectively extract meaningful information from high-dimensional analyses, a problem that is central in many fields. 

Most often, high-dimensional dataset extracted from, \textit{e.g.}, molecular dynamics (MD) trajectories (but this is true for any type of trajectory, including experimentally-resolved ones), are studied under an ergodic assumption as a collection of temporally-independent data, and clustering methods are used to detect high-density peaks identifying dominant/recurrent SOAP domains~\cite{Capelli2022,Gardin2022,Gasparotto2019}. 
However, we have recently shown that significant information may be hidden in the temporal correlations of the data -- such as local, sparse but dominant fluctuations, important local transitions, etc. -- that are typically overlooked in pattern recognition approaches which neglect time correlations. We have also developed an efficient method capable of maximizing the extraction and classification of information in time-series data, systematically distinguishing statistically relevant fluctuations from noise as a function of the time-resolution used in the analysis~\cite{Becchi2024layer}. 

In this work, we leverage such approaches to tackle fundamental questions relevant in high-dimensional analyses of molecular (and non-molecular) systems, such as: To what extent a high-dimensional analysis approach is really beneficial compared to lower-dimensional ones? How to compare information quantity \textit{vs.} information quality (relevance)? 
As a first prototypical example of an information-rich dataset, we analyze SOAP data extracted from atomistic MD simulation of a periodic box where liquid water and ice coexist in dynamical equilibrium at the melting temperature~\cite{Caruso2023tsoap, crippa2023machine, Becchi2024layer, Caruso2024, Martino2024}. This provides us with a prototypical system in equilibrium that respects a detailed balance and, at the same time, exhibits non-trivial internal complexity. This is explored here by computing 576-dimensional SOAP spectra for each of the 2048 water molecules in the system every $\delta t = 40$ ps (sampling interval) over 50~ns of MD trajectory. This results in a dataset of \(2.56 \times 10^6\) 576-dimensional SOAP spectra that we analyze both by ignoring or considering the time correlations between the data (i.e., pattern recognition {\it vs.} time-series analyses). 

Our results demonstrate how dimensionality reduction approaches, such as Principal Component Analysis (PCA), fall short when dealing with noisy data. Similar approaches allow discriminating and classifying only two solid ice and liquid water environments: \textit{i.e.}, the two most populated dominant phases in the system (recently Time-lagged Independent Component Analysis(tICA)~\cite{molgedey1994,perez2013} analyses were also proven not much more instructive).\cite{Caruso2024} 
On the other hand, we show how time-series data contained in a single SOAP component, which accounts remarkably only $<0.001\%$ of the total variance of the SOAP dataset (and that is thus typically discarded in PCA), can provide more information than that attainable by analyzing all dimensions in the SOAP dataset altogether. 
Our results reveal that adding more (noisy) components to this single one can be not only non-beneficial but also detrimental to the analysis, a phenomenon we describe as ``noise-frustrated information". Finally, we analyze a completely different system -- whose trajectories are obtained experimentally and are analyzed by combining physics-based descriptors -- showing that such effects are not exclusive of SOAP data, but can arise in any noisy multidimensional analysis. The findings presented here challenge classical paradigms in data analysis, highlighting the critical importance of focusing on information quality (relevance) rather than quantity to minimize information loss and misinterpretation.

\section*{Results}

\begin{figure}[htbp]
\centering
    \includegraphics[width=1\textwidth]{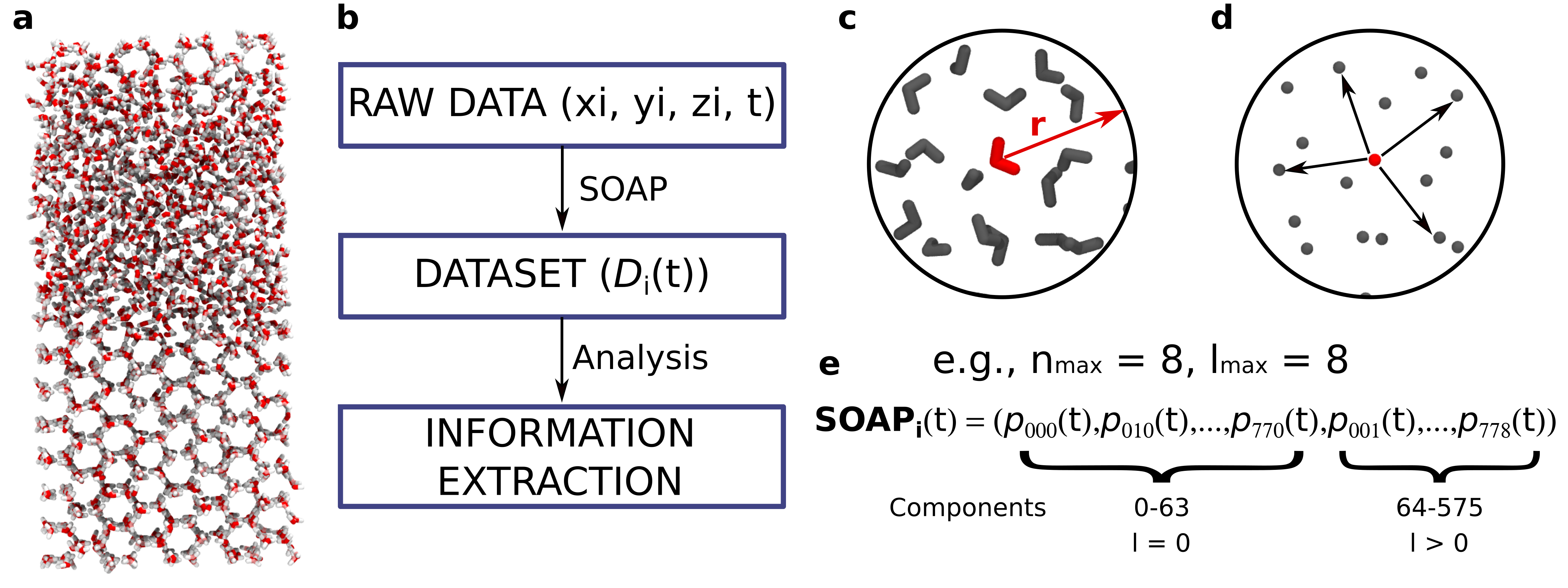}
    \caption{\textbf{Ice/water coexistence MD simulation.} a) Screenshot of the system studied. b) Schematic representation of the steps necessary to extract information from simulated systems. c) Zoom into water molecules, centered in a molecule (red) within a cutoff $r$. d) Zoom of the water molecules, considering exclusively oxygen atoms, which are used to calculate the SOAP descriptor. e) Definition of the Smooth Overlap of Atomic Position (SOAP) descriptor with distinction of components with $l = 0$ and $l > 0$.}
    \label{fig1}
\end{figure}

In this work, we use as a case study a 50~ns-long MD trajectory of an atomic system where liquid water and ice coexist in dynamic equilibrium (see Figure~\ref{fig1}a). Briefly, the periodic simulation box contains 2048 TIP4P/ICE water molecules~\cite{abascal2005potential}, initially arranged with 50\(\%\) in the liquid state and 50\(\%\) in a hexagonal ice configuration. The molecules are simulated in periodic boundary conditions at the melting temperature (see Methods section for details). This system has been recently proven to exhibit significant internal complexity~\cite{Caruso2023tsoap,crippa2023machine,Caruso2024,Martino2024}. 

The extraction of information from the trajectories of such a simulated system involves three main steps: (i) acquisition of raw data, which consist in the sequence of the particles' coordinates acquired every $40$~ps, over $\tau=50$~ns of MD simulation; (ii) selection of an appropriate descriptor $D$ to convert the trajectory of each molecule $i$ -- given by the time-series data \(x_i(t),y_i(t),z_i(t)\) -- into a dataset $\{D_i(t)\}$; (iii) analysis of the dataset to extract relevant information on the system's physics (see Figure~\ref{fig1}b). 
The information that can be effectively collected from the system depends on the choice of the descriptor in step (ii). The correct choice of descriptor is crucial for two main reasons. First, different descriptors capture distinct aspects of the system's behavior: for example, a structural descriptor would be useless to study purely dynamical features~\cite{crippa2023machine}. Second, each descriptor has an intrinsic signal-to-noise ratio, which influences the clarity and reliability of the extracted information~\cite{Martino2024}. When the physics of a system is unknown, selecting a general, agnostic -- possibly intrinsically high-dimensional -- descriptor would be desirable in order to minimize the risk of introducing assumptions that could bias the analysis or result in incomplete information. 

\subsection*{Building the high-dimensional (SOAP) dataset}  
Among the various descriptors used to study the coexistence of ice and liquid water~\cite{fitzner2019ice, crippa2023detecting, donkor2023machine, Caruso2023tsoap, Donkor2024spatavg, Martino2024}, here we employ SOAP~\cite{bartok2013representing} as a representative example of an abstract, general descriptor with an intrinsically high-dimensional nature. Recent studies have demonstrated SOAP's effectiveness in capturing complex phenomena, such as liquid-liquid and liquid-solid coexistence, early-phase transition nucleation, and other non-trivial processes~\cite{Capelli2022, Caruso2023tsoap, Donkor2024spatavg}. 
SOAP is a representation of the local particle density around each particle in the system through a power spectrum $\mathbf{p}$, whose components \(p_{nn'l}\) are indexed by radial ($n$ and $n'$) and angular ($l$) indexes. These components encode information about the local density and order, capturing features such as symmetry and the relative spatial arrangements of neighbors around each particle. This allows for information-rich analyses and enables the discrimination of different local environments within the system, evaluation of system homogeneity, and more. 

On a practical level, the results of SOAP analyses depend on parameters such as the maximum values $n_{\text{max}}$ and $l_{\text{max}}$. Selecting appropriate values for these parameters is crucial to balance the retention of information relevant to understanding the system's physics while avoiding an unnecessarily large number of components that could make the analysis computationally prohibitive. 
In the next section, we present results obtained using $n_{\text{max}} = l_{\text{max}} = 8$, the default value in the SOAP computation package \textit{DScribe}~\cite{himanen2020dscribe, laakso2023updates}. These values have been extensively used in prior works~\cite{Capelli2022, Caruso2023tsoap, crippa2023machine, Becchi2024layer, Caruso2024}. 
However, consistent results are also obtained using $n_{\text{max}} = l_{\text{max}} = 4$ (see results in the Supporting Figures~SI1-SI2). Using $n_{\text{max}} = l_{\text{max}} = 8$, we computed one 576-components SOAP power spectrum for each of the 2048 molecules in the system every \(\Delta t = 40\) ps over the course of \(\tau = 50\) ns of MD simulation (1250 frames in total). This resulted in a large high-dimensional dataset containing \(2.56 \times 10^6\) SOAP spectra, each with 576 components. 

\begin{figure*}[htbp]
\centering
    \includegraphics[width=0.98\textwidth]{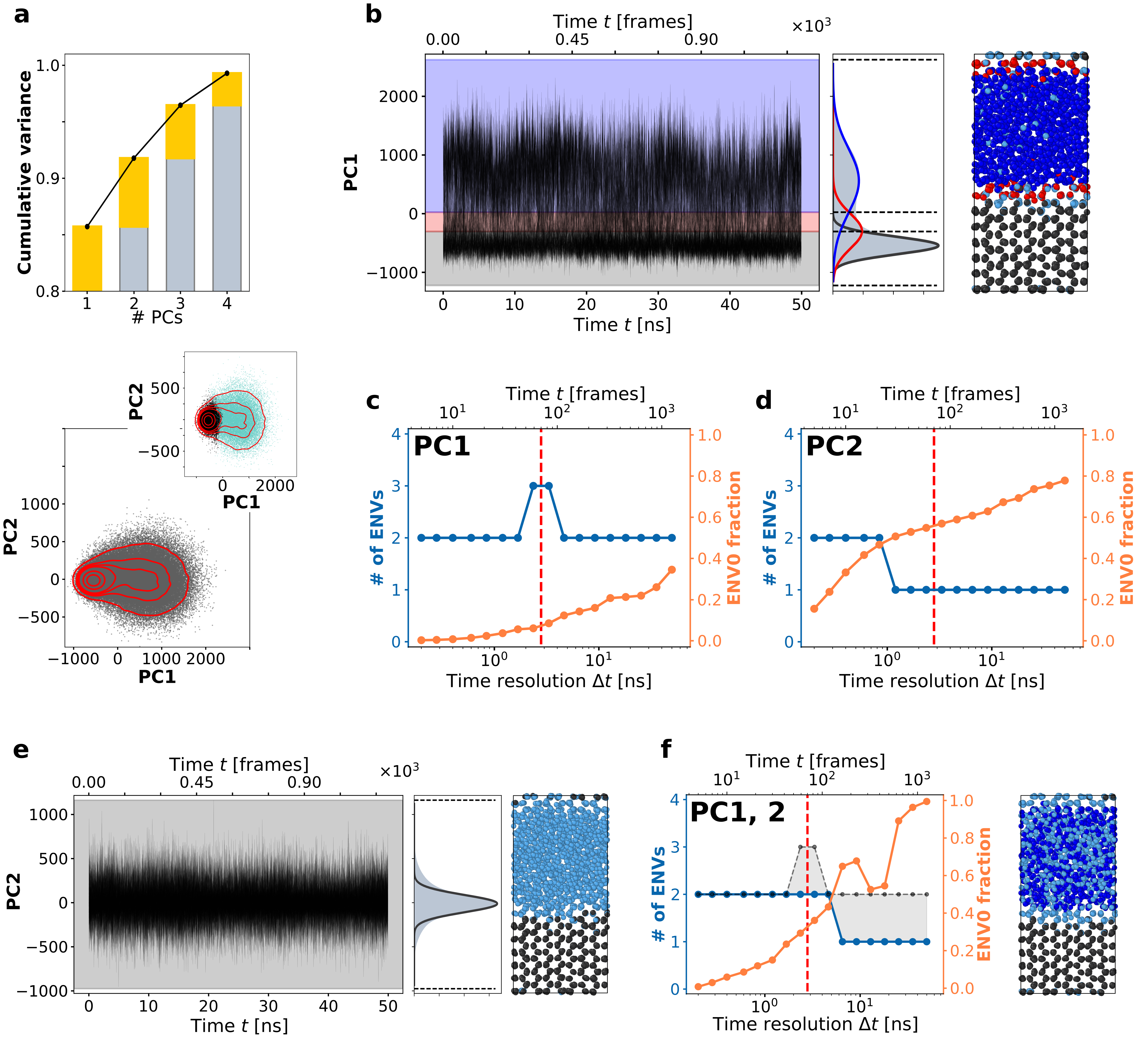}
    \caption{\textbf{The information nested along the time dimension: Onion Clustering on PC1 and PC2 time-series data.} a) Top: Cumulative data variance explained by the first four PCs. Bottom: SOAP dataset projection onto the first 2 PCs. Red contour isolines help to visualize the data density. Inset: common clustering approaches (hierarchical clustering) distinguish two main clusters (solid ice and liquid water) in the SOAP dataset. b) Left: PC1 time-series data of all the molecules. Center: kernel density estimate (KDE) of the PC1 time-series data, with the Gaussian environments (solid curves) identified by Onion Clustering~\cite{Becchi2024layer}. The dark gray environment corresponds to solid ice, the red one to the ice/water interface, and the blue one to liquid water. Right: snapshot of the MD simulation, where the water molecules are colored according to the Onion micro-clusters. c) Onion output plot, indicating the number of clusters (in blue) identified by Onion Clustering, and the fraction of lost information (in orange: unclassified information (ENV0) -- fast SOAP changes -- due to insufficient resolution) as a function of the time-resolution (\(\Delta t\)) used in the analysis of the PC1 time-series. d) Same as panel c), but for the PC2 time-series. e) Same as panel b), but for the PC2 time-series. f) Left: same results as panel c), obtained by Onion Clustering analyzing a bivariate (bi-dimensional) PC1, PC2 time-series (classified clusters in blue, unclassifiable information in orange). The gray dotted line shows the best result obtained by either of the two single components: the gray area shows the information lost by combining both dimensions (gray = max(\#$_{clust}$(PC1),\#$_{clust}$(PC2))-\#$_{clust}$(PC1,PC2)). Right: snapshot of the MD simulation, where the water molecules are colored according to the Onion micro-clusters detected at the resolution of \(\Delta t = 3\) ns (red dotted vertical line) from the bi-dimensional (PC1,PC2) time-series. }
    \label{fig2}
\end{figure*}

\subsection*{The information nested in the time-dimension} 
Common approaches to analyzing high-dimensional datasets require first to reduce their dimensionality. One of the most widely used techniques is Principal Component Analysis (PCA)~\cite{Pearson1901pca, Hotelling1936pca, Jolliffe2016pca}, based on the concept of variance. PCA reduces a high-dimensional dataset into a space of orthogonal features, ordering these components proportionally to the variance they explain. 
While PCA, like other dimensionality reduction methods, has intrinsic limitations (it works well only if the intrinsic feature space is linear), we use it here as an initial demonstrative test case to introduce fundamental concepts, which will be then explored using other approaches in subsequent sections. For the SOAP dataset derived from the water/ice trajectories analyzed in this study, the first four PCs (PC1-PC4) account for more than $99.3\%$ of the total variance (see Figure~\ref{fig2}a top). The projection of the full SOAP dataset onto the first two principal components, PC1 and PC2, is shown in Figure~\ref{fig2}a (bottom). In this projection, the areas of highest density (identified by the red isolines) correspond to recurrent and most-populated patterns -- or molecular motifs -- characterizing the system. 
The red contours reveal a density maximum at PC1 $\sim-500$ and PC2 $\sim0$, with a plateau extending to the right side of the plot. Using an unsupervised clustering method -- in this case hierarchical clustering~\cite{Nielsen2016hierarchical} -- we identified two domains: one encompassing the molecules in the density maximum and another including those in the plateau region (Figure~\ref{fig2}c, inset). Assigning the water molecules to these two clusters, based on this analysis, allows distinguishing the molecule in the solid (black) and in the liquid phase (cyan). 

It is worth noting that such pattern recognition approaches are based on an ergodic assumption, where temporal correlations between the data are ignored, and all these spectra are treated as part of a single static dataset. 
However, it is established that relevant information is contained in the time correlation of data. This has been demonstrated by various approaches, such as Markov state models and time-series analyses~\cite{gupta2013outlier, butler2024change}. Recently, we demonstrated that tracking the temporal sequence of individual signals from each molecule in a system can yield invaluable microscopic-level insights, which are essential for reconstructing both the microscopic and global behavior of the entire system~\cite{crippa2023detecting, Caruso2023tsoap, crippa2023machine, Empereurmot2023, Becchi2024layer, Cioni2024, Caruso2024, perrone2024}. 
In particular, we developed a fully unsupervised and essentially parameter-free clustering method, known as Onion Clustering~\cite{Becchi2024layer}, which can systematically detects statistically relevant fluctuations in noisy time-series data of any type. Briefly, the Onion Clustering identifies and classifies all dynamical environments within time-series data, proceeding with a hierarchical approach. Starting from the most evident features, the method iteratively detects, classifies, and archives them, continuing until no further statistically robust dynamical clusters can be identified~\cite{Becchi2024layer}. For a detailed description of the Onion Clustering method, we refer the interested readers to Ref.~\cite{Becchi2024layer}. 

In this work, we apply Onion Clustering to extract and classify the information contained in the SOAP dataset, which is studied as an ensemble of time-series rather than of uncorrelated SOAP spectra. 
We began by analyzing the time-series of the SOAP PC1 for all the 2048 molecules (see Figure~\ref{fig2}b (left)). Notably, PC1 alone accounts for \(\sim 86\%\) of the cumulative variance of the entire SOAP dataset, as shown in Figure~\ref{fig2}a. Onion Clustering successfully identified three statistically distinct clusters within the PC1 time-series, represented by the three Gaussian distributions shown in Figure~\ref{fig2} (center). These clusters correspond to three distinct environments, each characterized by a different average value of the PC1 SOAP spectra and a unique internal variance.  
These findings highlight that, although focusing solely on PC1 captures only a portion of the information contained in the full dataset, incorporating temporal correlations rather than treating the data as spatially and temporally de-correlated can yield more insightful results. 
Specifically, while the two main clusters -- solid ice and liquid water -- are detectable using standard pattern recognition methods applied to the full dataset (Figure~\ref{fig2}a, bottom), Onion Clustering additionally resolves a third cluster: the ice-water interface. 
This observation underscores an essential concept: the quantity of information in a dataset (its completeness) may be less critical than the quality (or relevance) of the information extracted. Here, the temporal correlations within the PC1 time-series contain more pertinent information than the entire SOAP dataset treated as static. This suggests that in certain cases, particularly for noisy datasets, ``less may be more" -- a principle that will be further demonstrated in the following sections. 

\subsection*{Information loss in time-series analyses: less may be more in time...}
Another example of the ``less may be more" principle becomes evident when considering how changes in time-resolution affect clustering results in time-series analyses. By design, Onion Clustering analyzes time-series data across all possible time-resolutions -- from the highest resolution (in this case, $\Delta t = 2\times40\text{ ps} = 80$~ps, corresponding to two trajectory frames), to the lowest resolution (here \(\Delta t = 50\)~ns, encompassing the entire simulation). The results for PC1 and PC2 are shown in Figure~\ref{fig2}c,d, while the results for PC3 and PC4 are reported in Supporting Figure~SI3. In Figure~\ref{fig2}c, the blue line shows the number of clusters detected as a function of time-resolution, while the orange line depicts the fraction of data points that can not be classified into a stable environment at the chosen resolution (ENV0), corresponding to molecules undergoing faster transitions. 

A key observation is that the third cluster, corresponding to the ice-water interface domain, can only be detected as a distinct environment within a specific time-resolution range \(2 \leq \Delta t \leq 4\) ns. The exact \(\Delta t\) window where this third environment can be resolved depends on both the system and the descriptor used~\cite{Martino2024}. However, in general, as time-resolution decreases (\(\Delta t > 4\) ns), the number of clusters reduces from three to two. This reduction occurs because the characteristic residence time of a molecule at the ice-water interface is at most $\sim4$~ns. When the analysis resolution falls below this threshold, the interface becomes indistinguishable from the other two environments, leading to the reassignment of interfacial molecules to the bulk ice or liquid environment. 
Conversely, using an excessively high time-resolution (\(\Delta t < 2\) ns) leads to ``oversampling" the time-series, saturating the dataset with information related to molecular vibrations. This oversampling reduces the statistical relevance of sparsely observed solid-to-liquid (and back) transitions. In this case, while sub-nanosecond molecular vibrations effectively distinguish solid ice from liquid water, they fail to resolve molecules at the ice/liquid interface. These interface molecules exhibit longer-scale dynamics and undergo sharp transitions on a nanosecond timescale, which are overlooked when the analysis focuses on finer temporal details. 

These findings demonstrate that increasing time-resolution in the Onion Clustering analysis does not necessary yield greater insights and higher-quality information. Instead, particularly when dominated by noise or irrelevant details, it can obscure important transitions, potentially leading to information loss. The results in Figure~\ref{fig2}a-c highlight the importance of the temporal component in uncovering relevant information. For this reason, we focus the study on the results obtained by Onion Clustering, which emerges as an optimal method for revealing the hidden dynamics within time-series datasets. 

\subsection*{...and in space: information loss in high-dimensional analyses }
The same approach, used to analyze the PC2 time-series data, leads to different results than those from PC1 (Figure~\ref{fig2}e). Unlike PC1, the PC2 time-series data exhibits higher noise levels and present a single density peak. Onion Clustering identifies only two environments in the PC2 time-series, corresponding to ice and liquid water. However, as the time-resolution of the analysis decreases, even at relatively high resolutions (\(\Delta t > 0.7\) ns), liquid water is not no longer detected as a statistically significant cluster and instead appears as a cluster of unclassified data (Figure~\ref{fig2}e, right). Notably, while PC2 accounts for a significant portion of the cumulative variance (\(\sim 6\%\)) of the dataset, it does not provide information about the ice-liquid water interface environment. Instead, it only retains information about the ice and liquid water environments, which were already prominent in PC1. This raises a fundamental question about the actual benefit of incorporating PC2 into the analysis when its contribution appears redundant compared to the insights already obtained from PC1. 

High-dimensional analyses are often thought as beneficial -- or even necessary -- for avoiding information loss. To explore this, we performed a bi-dimensional Onion Clustering analysis on a bi-variate PC1-PC2 time-series, combining information from both components. 
Theoretically, the combination of the PC1 and PC2 components should enhance the analysis by increasing the cumulative variance captured, from \(\sim 86\%\) to \(\sim 92\%\). However, as shown in Figure~\ref{fig2}f, the results reveal the opposite effect. 
The comparison between the blue line, which identifies at most two clusters at all resolutions, and the corresponding data for PC1 in Figure~\ref{fig2}c is particularly revealing. The black dashed line in Figure~\ref{fig2}f represent the minimum expected advantage when adding PC2 information to PC1. Conceptually, if adding dimensions were truly beneficial, the high-dimensional analysis (PC1, PC2) should never yield less information than the maximum achievable by using either dimension (PC1 or PC2) alone. In this case, since PC1 resolves more clusters than PC2 at every \(\Delta t\), the two-dimensional analysis should, at minimum, match the clustering results of PC1 alone (represented by the black dashed line in Figure~\ref{fig2}f). 
However, the results of Figure~\ref{fig2}f show that the bi-dimensional PC1-PC2 analysis fails to detect the interface environment altogether, with the number of clusters falling below the black line for all time-resolutions \(\Delta t > 1\) ns. 
This finding provides a concrete example of the ``curse of dimensionality"~\cite{friedman1997bias, bishop2006pattern}, where the introduction of additional spatial dimensions complicates the resolution of relevant information and ultimately leads to information loss. In the next section, we discuss the physical basis for such effects. 

\subsection*{Relevant information \textit{vs.} noise}

\begin{figure*} [htbp]
\centering
    \includegraphics[width=0.98\textwidth]{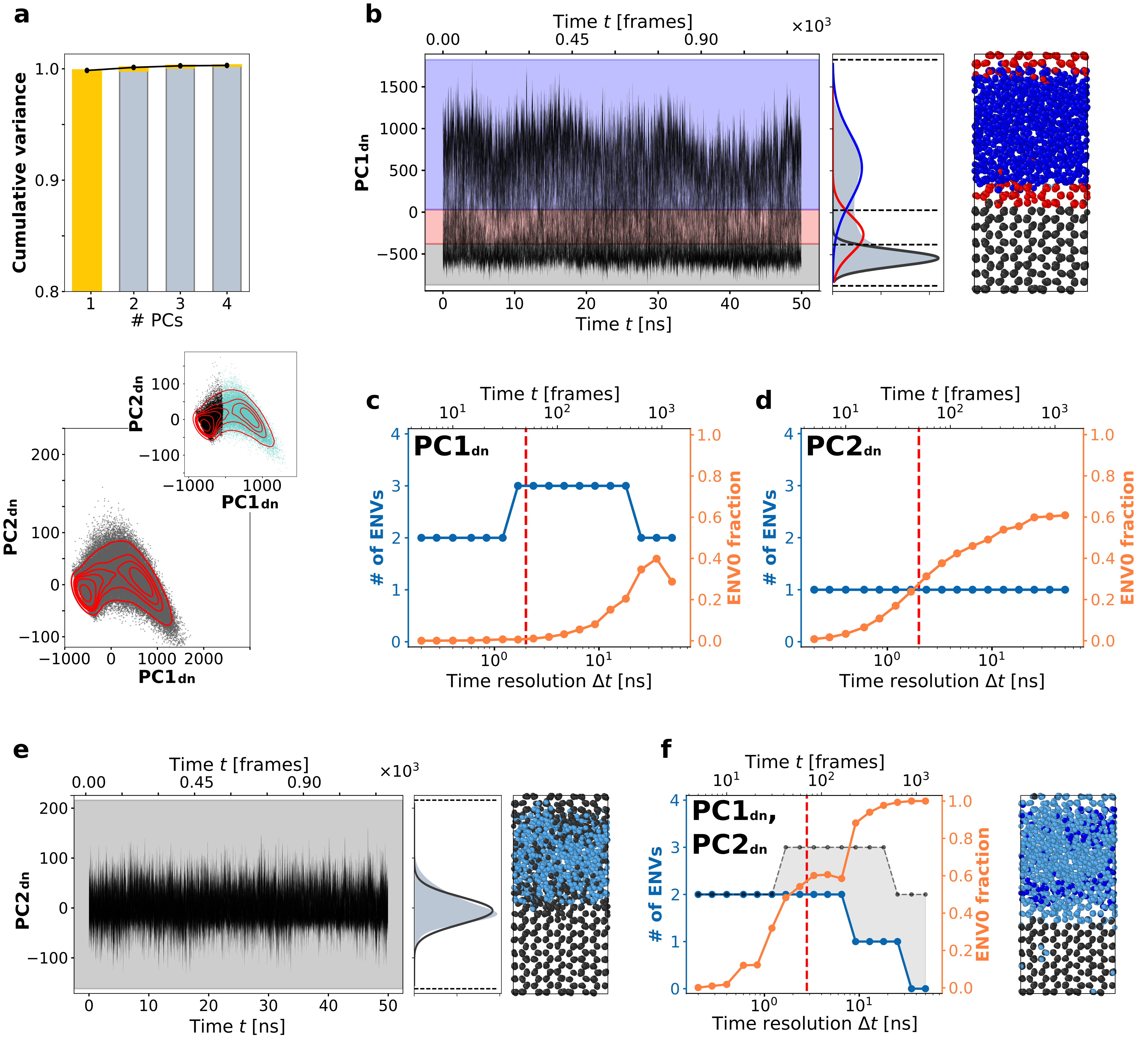}
    \caption{\textbf{Onion Clustering on denoised PC1 and PC2.} The figure follows the same structure of Figure~\ref{fig2}, but all the analysis are performed on the smoother time-series PC1$_{\text{dn}}$ and PC2$_{\text{dn}}$. The results of PC3$_{\text{dn}}$ and PC4$_{\text{dn}}$ are reported in Supporting Figure~SI4 }
    \label{fig3}
\end{figure*}

Descriptors like SOAP, rich in ``local information", can be significantly affected by local noise. Recently, Donkor {\it et al.} introduced an effective approach to reduce the local noise \textit{via} spatial average, thus by averaging the SOAP spectrum of each particle with those of its neighbors within the SOAP cutoff~\cite{Donkor2024spatavg}. The result is a denoised SOAP descriptor, with a decreased heterogeneity in SOAP spectra, enhancing the sensitivity of the subsequent analysis~\cite{Donkor2024spatavg, Martino2024}. 
Here, we illustrate the impact of this local denoising on our case study.
In our analyses, local-noise reduction in the principal components time-series produces notable effects. As shown in Figure~\ref{fig3}a (top), the cumulative variance of PC1 increases to \(\sim99.5\%\), compared to the original \(\sim86\%\) observed in Figure~\ref{fig2}a. This means that the other PCs contribute only marginal additional information (\textit{e.g.}, PC2 adds only \(\sim0.3\%\)). The projection of PC1 and PC2 in Figure~\ref{fig3}a (bottom) further demonstrate the effect of denoising (raw results are reported in Figure~\ref{fig2}a, bottom): two distinct density peaks, corresponding to the solid and liquid domains, are clearly visible. These clusters can be easily identified by any clustering method, as shown by the gray and light blue regions in the inset. 

It is particularly noteworthy that, in Figure~\ref{fig2}a, four PCs were required to reach \(\sim99.5\%\) of the cumulative variance, implying that the dataset was approximately four-dimensional. However, the local noise reduction allows PC1 alone to account for nearly the entire variance of the dataset, indicating that the ``relevant" information, without noise, is essentially mono-dimensional. This suggests that the high-dimensionality observed in the original dataset is primarily a reflection of noise, not of the inherent complexity of the system. 

We applied Onion Clustering to analyze the denoised PC1 (PC1$_{\text{dn}}$), which successfully identifies three distinct environments: ice, liquid water, and the liquid/solid interface (Figure~\ref{fig3}b). Compared to the results from Figure~\ref{fig2}b, the clusters in this case exhibit better separation, with the interface being more clearly detected and appearing thicker. Figure~\ref{fig3}c shows the number of clusters detected as a function of time-resolution. Notably, the interface environment becomes detectable at \(\Delta t = 1-2\)~ns, as observed in Figure~\ref{fig2}, but the reduction of local noise extends the interface's detectability up to \(\Delta t \sim 20\)~ns. approximately ten times more detectable in PC1$_{\text{dn}}$ than in PC1. This demonstrates the crucial role of local-noise reduction in time-series data, enabling a more stable and robust detection of physically relevant environments and events. In contrast, the analysis of denoised PC2 (PC2$_{\text{dn}}$) identifies only a single environment, corresponding to ice, while all the other molecules and data points are not classifiable. 

Combining PC1$_{\text{dn}}$ and PC2$_{\text{dn}}$ into a bi-dimensional time-series introduces similar issues as for the PC1, PC2 case. Also in this case, conducting a high-dimensional analysis not only fails to provide any advantage, but also results in information loss (note the extended gray area in Figure~\ref{fig3}f). This suggests that such high-dimensionality-induced information loss is not merely due to the noise of the individual time-series: in fact, the gray area of Figure~\ref{fig3}f (indicating information loss upon adding PC2$_{\text{dn}}$ to PC1$_{\text{dn}}$) is even larger than that in Figure~\ref{fig2}f. Instead, this highlights a critical issue intrinsic to high-dimensional analyses. Adding dimensions does not necessarily introduce relevant information; more often, it guarantees adding noise. In other words, the information in one dimension (PC2$_{\text{dn}}$) manifest as noise relative to PC1$_{\text{dn}}$, degrading the quality of the bi-dimensional analysis compared to analyzing the PC1 time-series alone. 

This evidence rises further question about the effectiveness of PCA as a dimensionality reduction technique. While PCA is widely used, it is known that its linearity of PCA may have limitations in handling high-dimensional datasets. Alternative dimensionality reduction techniques could be considered, each offering distinct advantages and drawbacks~\cite{Donkor2024spatavg,Tribello2019}. However, in this work we are not interested in the specific results of the SOAP analysis or in how to optimize it, but rather on specific concepts that typically emerge in high-dimensional analyses. For this reason, instead of testing different automatic dimensionality reduction approaches, we investigate the quantity, quality, and additivity of the information across individual SOAP components.

\subsection*{Information quantity \textit{vs.} information quality in noisy datasets} 

\begin{figure*} [htbp]
\centering
    \includegraphics[width=0.98\textwidth]{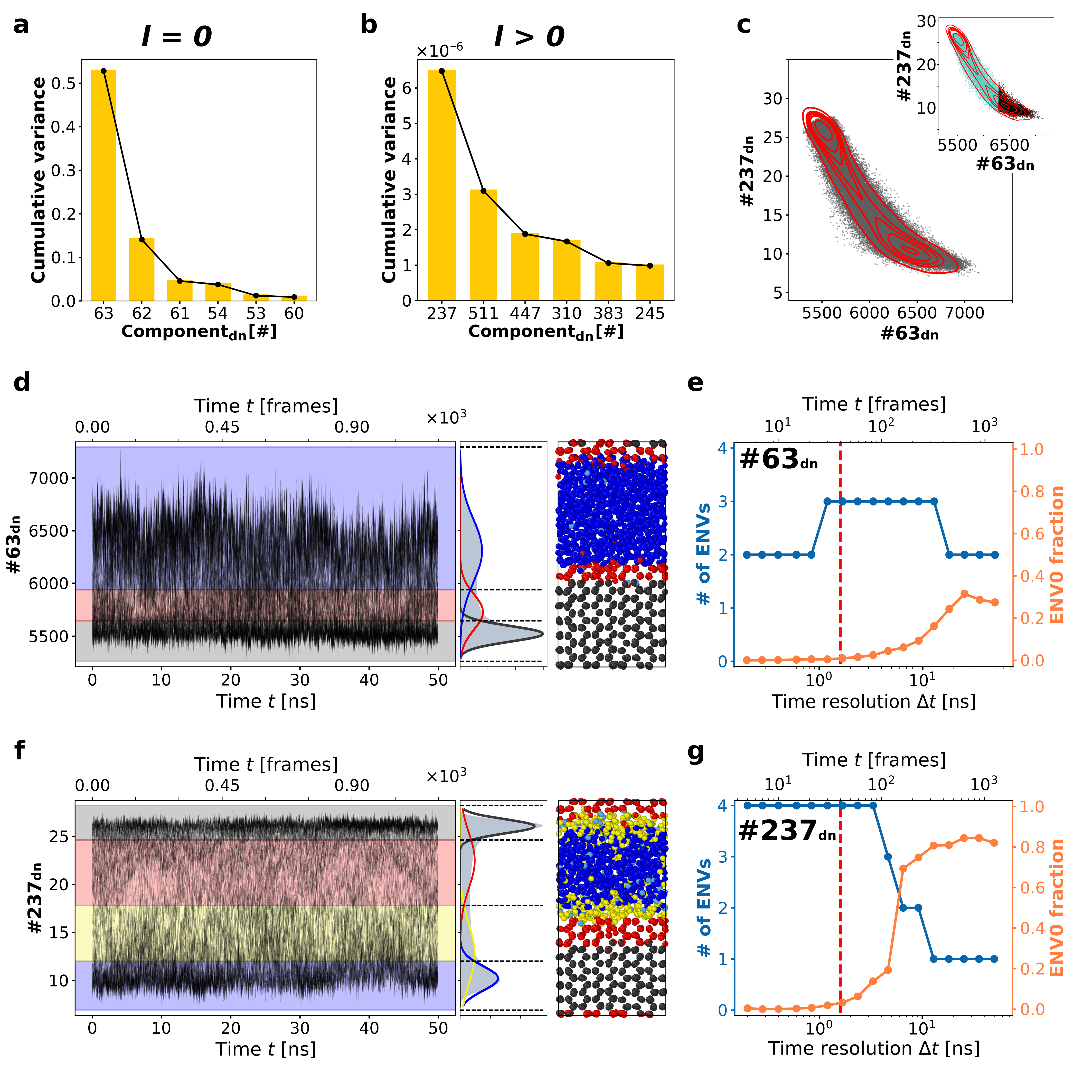}
    \caption{\textbf{Onion Clustering on denoised components.} a) Variance of the six most significant spherical ($l=0$) SOAP components. b) Variance of the six most significant non-spherical ($l>0$) SOAP components. c) Denoised dataset projection onto components \#63 and \#237; red contour lines help visualize the data density; in the inset, static clustering distinguishes 2 environments. d), e) Onion Clustering results of component $\#63_{\text{dn}}$. f), g) Onion Clustering results of component $\#237_{\text{dn}}$.}
    \label{fig4}
\end{figure*}

The intrinsic high-dimensionality of the SOAP spectra allow us to directly examine the information contained in its orthogonal \textit{n} components, without relying on any dimensionality reduction (thus avoiding the related issues). This is the primary reason why we used SOAP as a first example for this general study. 
PCA operated by selecting the PCs that maximize the explained variance, meaning, the amount (quantity) of captured information. In our analysis, the SOAP spectrum comprises 576 components, with approximately $53\%$ of the variance of the entire SOAP dataset captured by component $\#63_{\text{dn}}$ ($n=n'=8$,~$l=0$), as shown in Figure~\ref{fig4}b. The next most significant components, in terms of variance, $\#62_{\text{dn}}$ and $\#61_{\text{dn}}$, together with $\#63_{\text{dn}}$, account for  $\sim86\%$ of the total variance. 
It is already notable that for a system with such complex internal physics (typically assumed to require high-dimensional analysis~\cite{Capelli2022, Donkor2024spatavg, Caruso2023tsoap, Monserrat2020}) $\sim86\%$ of the cumulative variance is captured by just three SOAP components. Therefore, these three components are essentially contained within PC1$_{\text{dn}}$ (see Supporting Figure~SI5). Interestingly, components $\#63_{\text{dn}}$, $\#62_{\text{dn}}$, and $\#61_{\text{dn}}$ are spherical components ($l = 0$), meaning that they contain only information on the distances of the neighbors around every molecule and not on their orientations. 

The SOAP power spectrum becomes particularly interesting and enriched with relevant information for components with \(l > 0\). Unlike components with \(l = 0\), which lack of information about the relative orientation of neighboring molecules and encode only their distances, components with \(l > 0\) capture details about local symmetries and orientation. Figure~\ref{fig4}c highlights several \(l > 0\) components, ordered by their explained variance, starting from the highest. Notably, among them, component $\#237_{\text{dn}}$ is the one with the highest relevance: \(6\times10^{-6}\). 
In any variance-based dimensionality reduction approach, such as PCA, all \(l > 0\) components would likely be discarded due to their negligible statistical weight. However, while it might seem that these components contribute little to the understanding of the system's physics, this assumption is not necessarily accurate. 

We investigated the effects of combining the two components with the highest variance from the \(l = 0\) and \(l > 0\) families, specifically $\#63_{\text{dn}}$ and $\#237_{\text{dn}}$. By definition, these components are orthogonal in the SOAP space. In Figure~\ref{fig4}a, these two components are shown. Notably, an Onion Clustering analysis reveals that $\#63_{\text{dn}}$ alone (Figure~\ref{fig4}d) detects the interface more effectively than the PC1$_{\text{dn}}$ in previous cases (Figure~\ref{fig3}b). 
The increased separation between the peaks of the red, blue, and black Gaussian distributions demonstrates that the three different environments can be more easily discriminated in this case. As shown in Figure~\ref{fig4}e, $\#63_{\text{dn}}$ achieves comparable stability and robustness in detecting the third cluster to that of PC1$_{\text{dn}}$ in Figure~\ref{fig3}c, despite capturing only $\sim53\%$ of the variance compared to PC1$_{\text{dn}}$'s $\sim98\%$. This underscores the critical insight that variance, typically interpreted as the amount of information, does not necessarily correlate with the relevance or quality of that information. Furthermore, the fact that a single denoised component ($\#63_{\text{dn}}$) provides clearer and more discriminative information than PC1$_{\text{dn}}$ suggests that the combination of $\#63_{\text{dn}}$, $\#62_{\text{dn}}$, $\#61_{\text{dn}}$, and of the other components contributing to PC1$_{\text{dn}}$ increases the noise, thereby diminishing the clarity of the analysis. 

It is particularly fascinating to observe the results of Onion Clustering applied to the time-series of component $\#237_{\text{dn}}$, a variable with negligible variance (\(6\times10^{-6}\)) that would typically be overshadowed by the noise of the ``heavier" components. In this case, up to four distinct physically relevant environments are robustly detected at resolutions up to \(\Delta t \sim 3\)~ns. Specifically, the solid/liquid interface splits into two distinct layers: one representing liquid water in contact and exchange with ice, and the other representing ice in contact and exchange with liquid water, alongside the bulk of ice and liquid water phases. This result highlights how a single component ($\#237_{\text{dn}}$), despite it seemingly insignificant variance, can provide the most significant insights, underscoring the critical distinction between information quantity (variance) and information quality (relevance). 

\subsection*{The concept of  frustrated information} 

\begin{figure*} [ht]
\centering
    \includegraphics[width=1\textwidth]{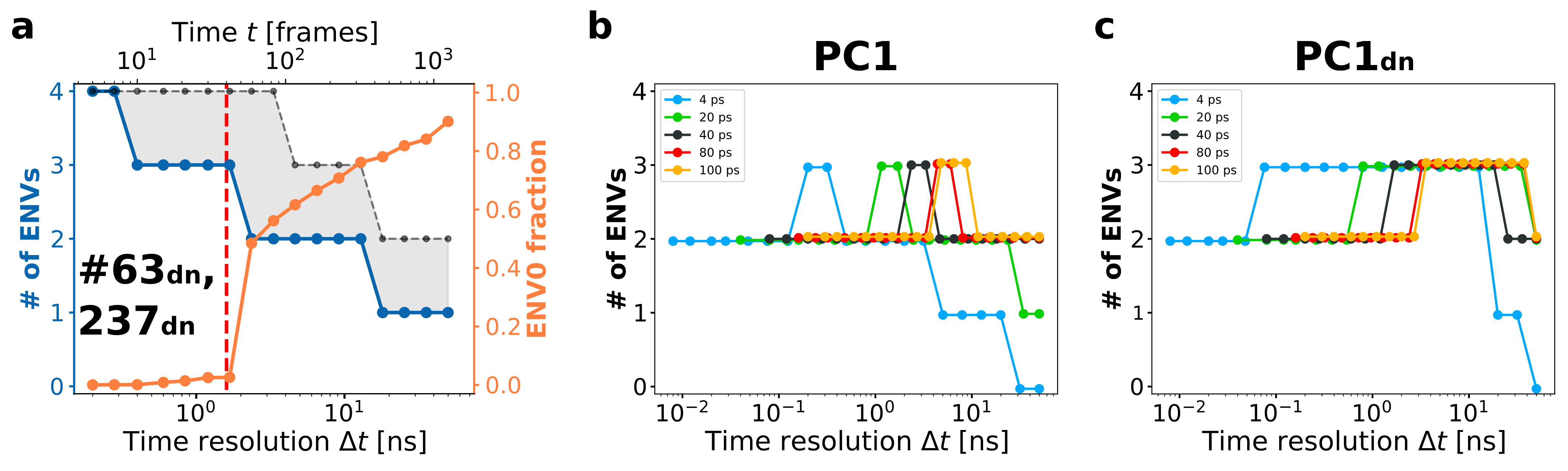}
    \caption{\textbf{Information frustration:} a) Onion Clustering on bivariate $\#63_{\text{dn}}$ and $\#237_{\text{dn}}$ time-series (see Figure~\ref{fig2}f for the legend). In gray, the combination of two components leads to considerable information loss, even when they are denoised. b-c) Dependence of the detection of the ice/water interface on the sampling \(\Delta t\) in the raw MD trajectories onto which the PC1 is calculated in the native/noisy (b) or denoised (c) SOAP dataset.}
    \label{fig5}
\end{figure*}

In principle, combining two orthogonal components, such as $\#63_{\text{dn}}$ and $\#237_{\text{dn}}$, should not result in signal degradation. At worst, their combination should preserve the maximum information obtainable from the individual components — specifically, detecting four clusters for \(\Delta t < 3\)~ns (guaranteed by $\#237_{\text{dn}}$), followed by a reduction to three and then two clusters at coarser resolutions. However, the bi-dimensional analysis still leads to significant information loss (Figure~\ref{fig5}a, gray area). 
Interestingly, this deterioration is not primarily due to superposition of noise from the two components (as shown in Supporting Figure~SI6). In fact, the information loss observed when combining the original, noisy, $\#63$ and $\#237$, components is smaller than that observed in Figure~\ref{fig5}a, which results from combining their denoised counterparts. Instead, this loss arises from an "information frustration" phenomenon, where the relevant information captured in one dimension ($\#63_{\text{dn}}$) appears as additional noise in the other ($\#237_{\text{dn}}$). 

Notably, all our evidences demonstrated that this phenomenon is general and can occur independently on the number of dimensions, their individual noise levels (native \textit{vs.} denoised), or the dimensionality reduction approach employed in the analysis. 
To further illustrate this point, Supporting Figure~SI7 shows the manifold representations of the SOAP and SOAP$_{\text{dn}}$ datasets (computed using UMAP~\cite{mcinnes2018umap, Donkor2024spatavg}). This approach operates directly on the full high-dimensional dataset -- meaning combining all component information ``correctly" -- and captures comparable physically relevant information to the one detected by the single component $\#237_{\text{dn}}$ (Figure~\ref{fig4}f, g). This confirms that discriminating the components by their variance is not necessarily the best approach, and better metrics should be used to extract the relevant information~\cite{Ebrahimi1999}. Similar to PCA, Time-lagged Independent Component Analysis (tICA)~\cite{molgedey1994,perez2013} performed on the same system, combined with KMeans clustering, also allowed to discriminate ice from water, struggling to resolve other dynamical microscopic domains (such as the ice-water interface).

Information frustration effects are also observed when additional data are added in time (by increasing the sampling frequencies) rather than in space (by adding SOAP components). To investigate the effects of different sampling frequencies, we compared the results obtained analyzing the MD trajectories (raw data) sampled every \(\Delta t = 40\)~ps (Figures \ref{fig2},\ref{fig3},\ref{fig4}) with those obtained by increasing or decreasing the sampling frequency. We therefore resampled the 50~ns MD trajectory at intervals of 4~ps, 20~ps, 80~ps, and 100~ps. 
The expectation is that Onion Clustering, which extracts relevant information about physical events in the system (for instance, the detectability of the ice/water interface environment up to a certain \(\Delta t\)), should maintain or enhance its ability to resolve this environment when using finer sampling (\(\Delta t < 40\)~ps). Conversely, coarser sampling (\(\Delta t > 40\)~ps) might compromise the resolution,  potentially preventing proper detection of the interface.
However, the results of these analyses reveal issues analogous to those encountered when adding more spatial information. As shown in Figure~\ref{fig5}b, the black line represents the reference case (\(\Delta t < 40\)~ps): the third environment -- the ice/water interface -- is detectable within the resolution range 2~ns $< \Delta t < 4$~ns in the PC1 time-series, as seen in Figure~\ref{fig2}c. Increasing the sampling time to 4~ps or 20~ps produces striking effects, as the temporal window within which the interface is detected shifts backward in the time-series. For a 20~ps sampling interval (which doubles the number of frames compared to the reference), the detectable temporal window for the interface shifts to 1~ns $< \Delta t < 2$~ns (green data). With a 4~ps sampling interval (a tenfold increase in the number of analyzed frames), the interface is detectable only between 200~ps and 300~ps, a much shorter period compared to the nanosecond-scale range observed in other cases. 

Remarkably, increasing the sampling frequency not only shifts the time window for detecting the same environment backward, but also makes its detection progressively less robust. For instance, doubling the analyzed data ($\Delta t = 20$~ps) reduces the detection robustness by 50$\%$ (from 2~ns with sampling $\Delta t = 40$~ps to 1~ns with $\Delta t = 20$~ps, and twenty times less robust when analyzing tenfold data (100~ps with $\Delta t = 4$~ps). This is a non-trivial outcome. Physically, the environment characterized by molecular exchange between ice and liquid water remains the same in all cases. However, the addition of temporal data highlights a significant dependency on the chosen sampling interval. One might intuitively expect that increasing the sampling frequency would enhance resolution and simply allow earlier detection of the interface. 
Instead, oversampling leads to unexpected consequences: not only does it diminish the robustness of detecting the event, but it also introduces ``data-drive hallucination", where the same physical event appears to exhibit different dynamics depending on the time-resolution. This phenomenon may suggest that the analysis itself is altering the reconstructed physics of the system. The addition of more data can generate an artifact, which appears to be an effect of oversampling and of the system's local noise. 

We repeated the same analysis on the denoised PC1s (Figure~\ref{fig5}c). The reduction of local noise supports such hypothesis: as the sampling interval increases, the detection of the interface shifts earlier, but the final time-resolution at which detection concludes remains consistent. In this case, the detection of the interface is more robust across all sampling frequencies, with an average detection range of 0.8~ns $\leq \Delta t \leq 40$~ns. This suggests that increasing the sampling frequency improves the ability to detect this dynamic environment sooner, while the lowest resolution capable of its detection remains consistent across different samplings. This behavior is governed by the inherent physics of the system, particularly the longest residence time of the molecules at the ice/water interface. 
Notably, the unique offset occurs in the case of a 4~ps sampling case (cyan). Here, even in the absence of noise, the interface is detectable only within a time window of $\sim10$~ns, compared to the broader $\sim40$~ns observed in the other cases. This provides a clear example of how oversampling, even with denoised data, can lead to information loss. 

The data in Figures~\ref{fig5}b,~c, together with those in Figure~\ref{fig5}a and of Figures~\ref{fig2},~\ref{fig3}, and~\ref{fig4} highlight a general take-home message: adding more data increases the total amount of information, which can be understood as the sum of relevant information and noise. However, while adding more data does not guarantee an increase in relevant information, it invariably introduces additional noise, potentially leading to oversampling and to information loss. In principle, increasing the amount of information is beneficial only if it is proven \textit{a priori} that the added data contain relevant information -- a determination that requires preliminary analysis. 
It is important to recognize that these issues are particularly inherent in typical pattern recognition approaches applied to ``static datasets", where the temporal information is lost. Furthermore, these phenomena are not exclusive to specific descriptors or to the SOAP dataset. On the opposite, as demonstrated in the next section, the effect of information frustration and information loss can arise (to a smaller or larger extent) whenever two or more features are combined in an attempt to enrich the analysis. 

\subsection*{Generality of the concept: Quincke rollers as a second different case study}

\begin{figure*} [htbp]
\centering
    \includegraphics[width=1\textwidth]{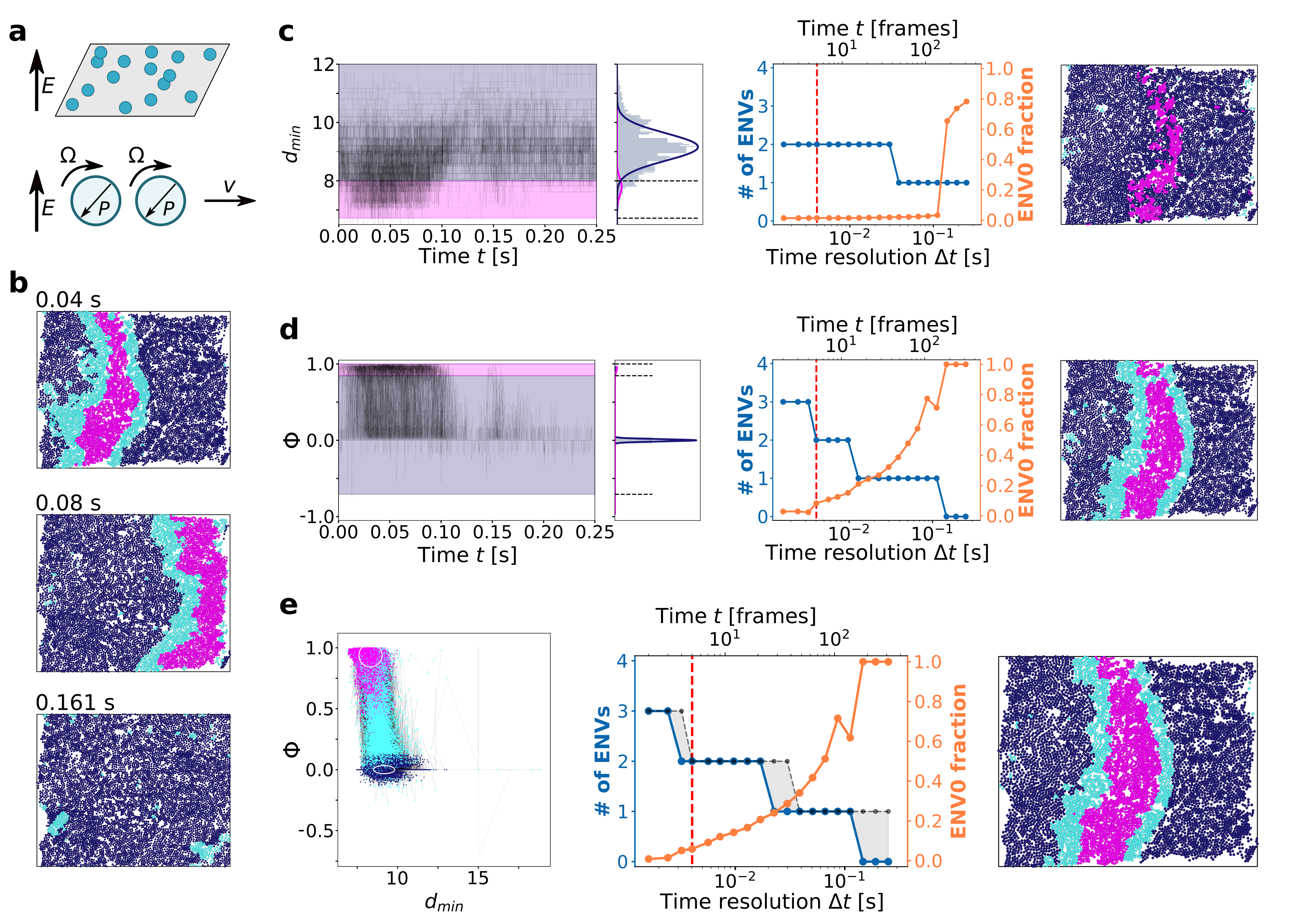}
    \caption{\textbf{Onion Clustering on experimental trajectories of dynamical Quincke rollers.} a) Schematic representation of Quincke rollers. When suspended in conductive fluid and exposed to a vertical electric field, the particles exhibit collective motions~\cite{Liu2021quinckerollers}. b) Three snapshots from the video showing the wave moving from left to right, colored according to clusters detected by Onion Clustering on $\Phi$ c) Onion Clustering output for spatial averaged $d_{min}$. d) Same as panel c), but for denoised $\Phi$. e) Same as in panel c), but for the clustering on both spatial averaged $d_{min}$ and $\Phi$. }
    \label{fig6}
\end{figure*}

In the tests of Figures~\ref{fig1}-\ref{fig5} we used SOAP as a representative example of a high-dimensional descriptor, providing high-dimensional datasets. However, the same phenomena can occur independently of the specific system or the set of descriptors/variables selected.
To illustrate the generality of these concepts, we analyzed a completely different system, characterized by unique complex internal dynamics. In this case, we examined time-series obtained from experimentally-acquired trajectories of a system involving the two-dimensional collective motions of colloidal polymeric microparticles, known as Quincke rollers. Readers seeking more details on this specific system are referred to Refs.~\cite{Liu2021quinckerollers, Becchi2024layer, Caruso2024}. 
Briefly, Quincke rollers are dielectric polystyrene particles ($\sim9.9~\mu$m in diameter) suspended in a weakly conductive fluid and confined within a two-dimensional cell. Under the influence of a weak vertical DC electric field, these particles can exhibit complex collective motions, such as vortexes and waves. 
While this is a higher-scale system, completely different from the ice-water molecular-scale one studied in the previous sections, it is useful to show how also in this case we encounter similar "information frustration" effects/issues, which are thus not limited to certain scales, systems, or descriptors, but are general in high-dimensional analyses of noisy systems.

Here, we analyze the trajectory we resolved for one example Quincke rollers system~\cite{Becchi2024layer}, consisting of 6921 particles moving within a $700 \times 700~\mu$m cell over a real-time duration of 0.25~s (310~frames in total). During this interval, a collective density wave is seen to propagate from left to right in the microscopy field (see Figure~\ref{fig6}a-b). 

For the sake of generality, in this case our analysis uses two human-based descriptors. Recently, we have demonstrated that integrating static descriptors, such as SOAP, with dynamic ones, such as LENS, can enrich high-dimensional analyses~\cite{crippa2023detecting}. In the same spirit -- and reflecting common practice in studying collective dynamical events in active matter~\cite{peruani2008mean} -- we selected two descriptors: (i) the minimum neighbor distance $d_{\text{min}}$, as a proxy for the local particle density, and (ii) the local alignment of the particle velocity $\phi$. This choice results in a bi-dimensional analysis (details are provided in the Methods section). We analyzed the denoised mono- and bi-dimensional time-series of the 6921 particles. Using Onion Clustering, $d_{\text{min}}$ was identified as a less informative descriptor (Figure~\ref{fig6}c). From its time-series, two environments are detected up to $\Delta t \leq 30$~ms: one representing the wave's core (having higher density) and the other identifying static particles (with lower density). 
In contrast, the particles' local velocity alignment $\phi$ (Figure~\ref{fig6}d) is found to be more informative, capturing the system's collective motion and identifying three distinct environments for resolutions higher than $\Delta t \leq 3$~ms, and two environments for $\Delta t \leq 10$~ms. When three clusters are detected, Onion Clustering distinguishes the wave's core (Figure~\ref{fig6}d in fuchsia) and two regions (cyan) where particles transition between static and dynamic states around the wave’s core. These regions are shown in the snapshot in Figure~\ref{fig6}d, alongside with the static particles' environment (dark violet). 
For $\Delta t \geq 140$~ms, Onion Clustering fails to resolve any statistically distinct dynamical cluster, indicating that, in the system, all particles change environment at list once during the trajectory. The residence time of particles in any cluster is therefore less than 140~ms. 

While the variable $\phi$ is found more descriptive than $d_{\text{min}}$ for this system, it is interesting to examine the results of combining the two variables in a two-dimensional ($\phi$, $d_{\text{min}}$) time-series analysis. The outcomes are shown in Figure~\ref{fig6}e. Comparing the number of clusters identified in the bi-dimensional analysis to those from the individual mono-dimensional analyses confirms that $\phi$ captures all relevant information, while $d_{\text{min}}$ contributes minimally. From this perspective, one might expect the two-dimensional analysis to at least match the results obtained from the single-variable analysis of $\phi$. 
However, the number of clusters detected in the two-dimensional analysis is diminished compared to the one-dimensional clustering of $\phi$, as highlighted by the black line and gray area in Figure~\ref{fig6}e, center. Thus, the addition of a second variable ($d_{\text{min}}$) does not enrich the analysis, but instead leads to a loss of information. 
While the severity of the issue is somehow mitigated in this case -- likely due to the combination of position-dependent ($d_{\text{min}}$) and velocity-dependent ($\phi$) information~\cite{crippa2023machine}) -- this demonstrates that combining multiple descriptors can lead to non-trivial phenomena of information frustration and loss. This can occur even in relatively simple cases where the two variables are markedly different. 
Notably, this problem is not exclusive to specific systems or combinations of descriptors. Rather, it is a general issue in (high-dimensional) analysis, stemming from the fact that information overloading -- even when the individual pieces of information are intrinsically relevant -- can hinder the ability to discern meaningful details within a dataset. 

\section*{Conclusions}

Understanding the physics of complex systems is inherently challenging. It is often assumed that high-dimensional analyses are not preferable, but necessary to comprehensively capture the underlining physics of such systems. However, the actual necessity of high-dimensionality, as well as the methods to extract meaningful physical information from such data, often remain unclear. In this work, we address this fundamental issue by systematically exploring the use of high-dimensional abstract descriptors and combinations of multiple physics-inspired ones. Our findings reveal nontrivial and counterintuitive aspects of high-dimensional analyses, demonstrating that high-dimensionality is not always essential and, in some cases, can lead to frustrated information and even information loss.

As a prototypical case study, we analyzed Smooth Overlap of Atomic Position (SOAP) spectra~\cite{bartok2013representing} extracted from MD simulations of coexisting ice and liquid water at dynamic equilibrium at the solid/liquid transition temperature. Typically, SOAP analyses of such systems involve dimensionality reduction methods, such as Principal Component Analysis (PCA) followed by clustering based on a combination of principal components (PCs) that account for high cumulative variance. However, approaches like this ignore the time-correlation in the data; this can result in significant information loss and limit the ability to capture key dynamic events~\cite{Becchi2024layer}. 
To maximize the information extraction, we opted to analyze the SOAP data in time-series using Onion Clustering~\cite{Becchi2024layer}, which allows resolving the information within individual SOAP PC1 and PC2 time-series, as well as their combination in bi-variate time-series. Our results demonstrate that two-dimensional analysis systematically lead to information loss, independently of whether the time-series data are noisy or denoised~\cite{Donkor2024spatavg}. This phenomenon persists even when analyzing two of the original SOAP components, such as $\#$63 and $\#$237 (see Figures~\ref{fig4}-\ref{fig5}), and thus regardless of the chosen dimensionality reduction approach. We call this effect ``frustrated information", wherein relevant information captured in one dimension manifest as noise in another one. 

To ascertain what it is typically meant with ``relevant information", we systematically analyzed the individual SOAP components, separating them into radial ($l=0$) and angular ($l<0$) contributions. Remarkably, our findings reveal that a specific angular component -- $\#$237 -- which accounts for mere $10^{-6}$ in the total variance (and is typically overlooked in variance-based dimensionality reduction approaches), contains more relevant and discernible information than all other components, or even the entire dataset.
This shifts the focus from traditional ``dimensionality reduction" to a more nuanced ``data mining" perspective, highlighting that identifying the component or dimension containing the most relevant information is far more critical than simply prioritizing those with higher variance or combining multiple dimensions. 

Distinguishing between the information contained in temporal \textit{vs.} spatial dimensions reveals that adding more information than necessary in space (high-dimensionality) can lead to ``frustrated information". Similarly, excessive information in time (oversampling) can result in ``information hallucinations", where the same physical event appears to have different features depending on the analysis (Figure~\ref{fig5}). The varying timescales over which the ice/water interface -- an environment characterized by water molecules undergoing solid-to-liquid and reverse phase transition -- is detected in Figure~\ref{fig5}b,~c, clearly  illustrate the challenges of analyzing noisy data. These findings underscore how the observed physics of a phenomenon can appear to shift dramatically based on the chosen sampling or measurement method. 

While we used SOAP as a prototypical descriptor enabling high-dimensional analysis, the results shown in Figure~\ref{fig6}, obtained for a system with entirely different scale, dynamics, and descriptors, demonstrate that such issues of nested, hidden, and frustrated information can emerge in virtually any type of multidimensional analysis. These findings provide clear examples of the difference existing between information quantity and quality (relevance), showing that relying solely on variance-based approaches (which prioritize quantity) can be very risky -- especially in noisy, high-dimensional dataset, where the noise from one dimension may outweigh the relevant information in another one. At the same time, this work offers useful insights for navigating such complexities and extracting meaningful information from noisy dataset more effectively. 

\setcounter{section}{1}
{\small
\section*{Methods}

\subsection{Smooth Overlap of Atomic Position (SOAP)}

In this work, we use SOAP~\cite{bartok2013representing} as a prototypical example of high-dimensional descriptor, which offers a very high-dimensional representation of the local structural environment of each particle. In particular, here, the SOAP power spectrum is calculated for each oxygen atom at every time step $t$, using the \textit{Dynsight} software, available at~\cite{dynsight}, considering all other oxygen atoms within a cutoff radius of $r_{\text{cut}} = 10$~\AA~(previously demonstrated to be a good cutoff distance for aqueous systems~\cite{Caruso2023tsoap, Caruso2024, Capelli2022, Becchi2024layer, crippa2023detecting, crippa2023machine}. 
The SOAP vectors are computed using the \textit{DScribe} Python package~\cite{himanen2020dscribe}, using both $l_{\text{max}} = n_{\text{max}} = 8$ and $l_{\text{max}} = n_{\text{max}} = 4$, for comparison. In the first case, we obtain 576 components for the spectrum of each oxygen atom at each time step, while in the second case, the number of components is reduced to 80, globally obtaining in both cases consistent results (see Supporting Figures~SI1-SI2). The PCA to reduce the dimensionality of the dataset (Figures~\ref{fig2},~\ref{fig3}) were performed using the \textit{SciPy} Python package~\cite{virtanen2020scipy}.

\subsection{Simulations and data analysis}
\paragraph{Water/ice MD simulation:}
The atomistic system studied herein consists of 2048 TIP4P/ice molecules~\cite{abascal2005potential} initially set 50$\%$ in $Ih$ ice and 50$\%$ in liquid water phase, that we used to simulate the coexistence of hexagonal ice $Ih$ and liquid water~\cite{Matsumoto2017}. To maintain phase coexistence, MD simulations are performed at the melting temperature of $268 K$ in the TIP4P/ice model~\cite{abascal2005potential}. After an initial equilibration phase, production is carried out under $NPT$ and semi-isotropic conditions for 50~ns, with trajectory data sampled every $1$~ps, using the GROMACS software~\cite{Abraham2015}. The trajectory is then resampled based on the chosen sampling intervals (\textit{i.e.}, $4$~ps, $20$~ps, $40$~ps, $80$~ps, and $100$~ps, see Figure~\ref{fig5}b,~c). Further details on the MD simulation can also be found in Refs. \cite{crippa2023detecting, Caruso2023tsoap}. 

\paragraph{Quincke rollers trajectory and analysis}
The trajectory studied in Figure~\ref{fig6} is obtained from microscopy videos from  Ref.~\cite{Liu2021quinckerollers}. The position of each particle at every frame are extracted by applying an image recognition and particle tracking, performed with \textit{Trackpy}~\cite{trackpy}. The raw trajectory consists of 6921 particles in a $700 \times 700~\mu$m bi-dimensional cell over a real-time duration of 0.25~s sampled every $\Delta t = 0.8$~ms, for a total of 310~frames. For each molecule, we calculate the distance from the closest neighbor $d_{\text{min}}$, and the local alignment of the velocities $\phi$ as in Eq.~\ref{eq1}: 

\begin{equation} 
\phi_i = \frac{1}{n_c^i} \sum_j \frac{\mathbf{v}_i \cdot \mathbf{v}_j}{|\mathbf{v}_i| |\mathbf{v}_j|}
\label{eq1}
\end{equation}

where $j$ iterates over the $n_c^i$ particles located within a specific cutoff distance of $r_c = 15$ pixels from particle $i$~\cite{Becchi2024layer}. $\mathbf{v}_i$ and $\mathbf{v}_j$ are the velocities of particles $i$ and $j$, respectively. The variable $\phi_i$ represents the average cosine of the angle between the velocity of particle $i$ and those of its neighboring particles at every instant $t$. This value ranges from -1 to 1, reflecting the degree of alignment or orientation similarity between the velocities of the particle and its neighbors. Further details can also be found in Ref.~\cite{Becchi2024layer}.

\subsection{Time-series analysis with Onion Clustering}

Onion Clustering is a recently developed algorithm for single-point time-series clustering, and for robust detection and classification of statistically-relevant fluctuations in noisy time-series~\cite{Becchi2024layer}. Considering a time-series of length $\tau$ composed of \textit{n} consecutive frames with $\Delta \tau$ interval between each frame ($\tau = n\times \Delta \tau$). Onion Clustering conducts multiple clustering analyses at all possible resolutions -- from the maximum one, $\Delta t = 2\Delta \tau$, to the minimum one, $\Delta t = \tau$. At each resolution, Onion Clustering can classify the dynamically-persistent environments in which a system can be subdivided and the transitions between them, leveraging an iterative detect-classify-archive approach allowing to uncover all possible classifiable information at a given $\Delta t$. This allows to iteratively ``peel" the dynamics complexity of a time-series, proceeding with the identification of such microscopic dynamic domains from the most populated (dominant and evident) to the most hidden and least populated one, and classifying the molecules or units belonging to them at each $\Delta t$. In this work, to filter out statistically insignificant clusters, populations below $1\%$ are removed and labeled as unclassifiable information. 

We just underline that performing the analysis at varying time-resolutions ($\Delta t$) helps to automatically identify the optimal $\Delta t$ that maximizes the subdivision of the time-series (and the system) into distinct statistically-relevant microscopic environments and minimizes the fraction of unclassifiable points (ENV0). For further details, we refer the readers to Ref.~\cite{Becchi2024layer} for details on the onion method, and to Refs.~\cite{Martino2024, Caruso2024} for other recent applications. The Onion Clustering code is available on the platform \textit{Dynsight}~\cite{dynsight}.
}

\subsection{Data availability}

Complete data and materials pertaining to the trajectories employed, and the analysis conducted herein, are available at: \href{https://zenodo.org/records/14529457}{https://zenodo.org/records/14529457}. Other information needed is available from the corresponding author upon reasonable request.

\section*{Acknowledgments}

G.M.P. acknowledges the support received by the European Research Council (ERC) under the European Union’s Horizon 2020 research and innovation program (Grant Agreement no. 818776- DYNAPOL). G.M.P. and C.L. also acknowledge the support received by the ICSC—Centro Nazionale di Ricerca in High Performance Computing, Big Data and Quantum Computing, funded by European Union - NextGenerationEU.

\printbibliography

\setcounter{section}{0}
\setcounter{subsection}{0}
\setcounter{figure}{0}
\renewcommand{\thefigure}{SI\arabic{figure}}

\section*{Supporting Information}

\begin{figure}[H]
\centering
    \includegraphics[width=1\textwidth]{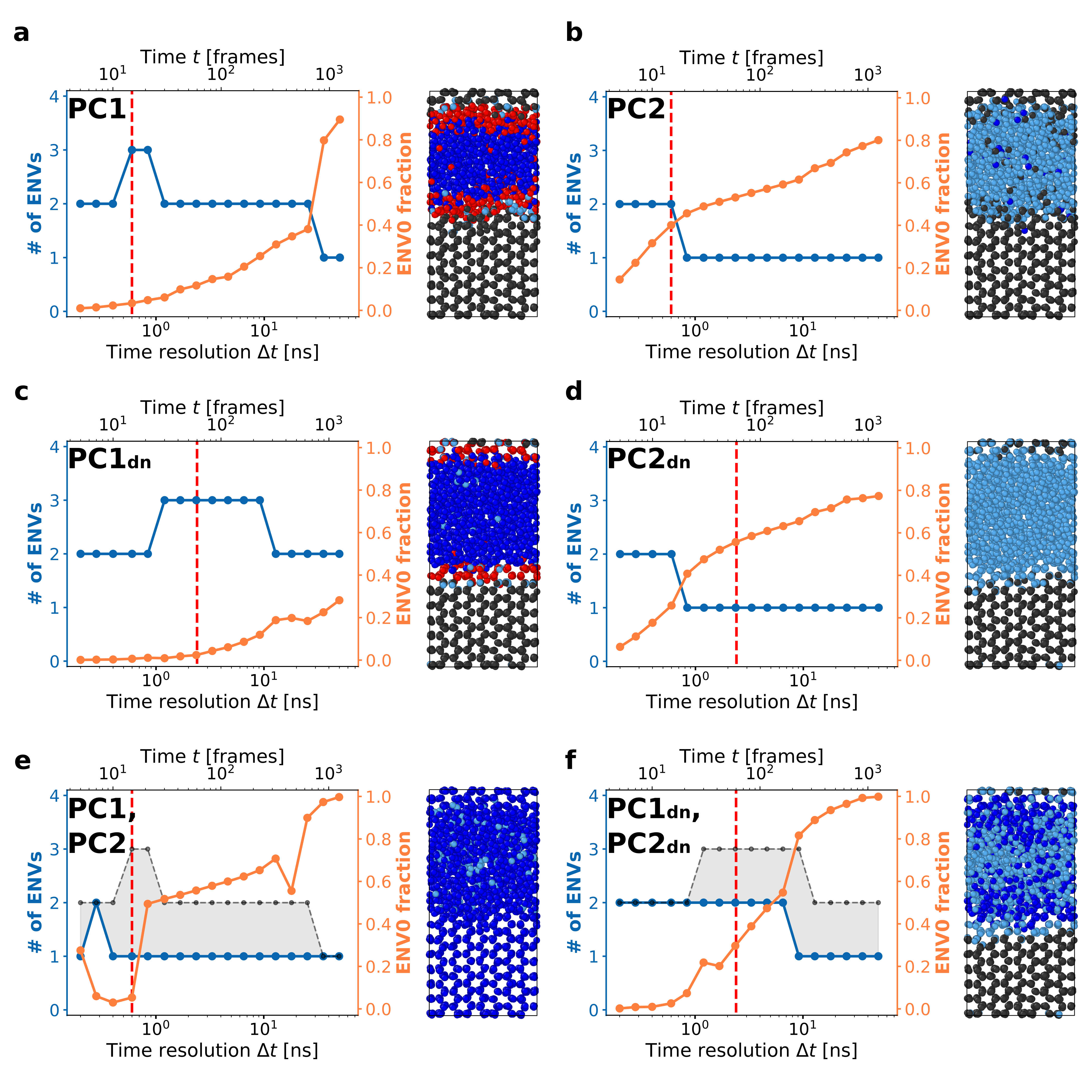}
    \caption{\textbf{Onion Clustering on PC1 and PC2 time-series data, obtained from SOAP with \(l_{max} = n_{max} = 4\).} a) and b) Onion Clustering results calculated for raw PC1 and PC2, and relative screenshots. c) and d) Onion Clustering results on PC1 and PC2 denoised, and relative screenshots. e) and f) Results from the Onion Clustering algorithm applied on the bi-dimensional (PC1, PC2) time-series, e) raw and f) denoised.}
    \label{figSI1}
\end{figure}

\begin{figure}[H]
\centering
    \includegraphics[width=1\textwidth]{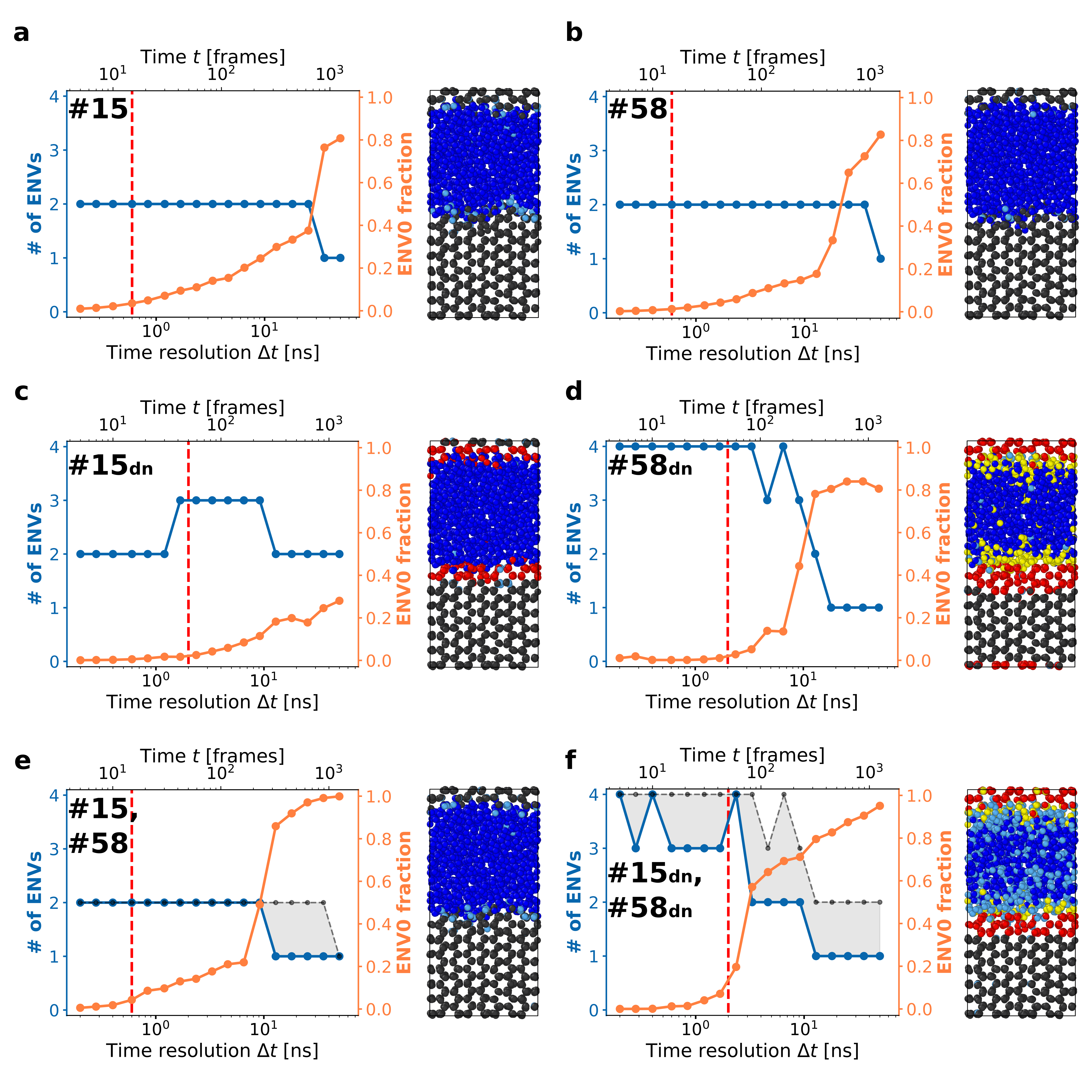}
    \caption{\textbf{Onion Clustering on components \(\#15\) (highest variance for \(l = 0\)) and \(\#58\) (highest variance for \(l > 0\)), both raw and denoised.} The figure follows the same structure of Figure~\ref{figSI1}. }
    \label{figSI2}
\end{figure}

\begin{figure}[H]
\centering
    \includegraphics[width=1\textwidth]{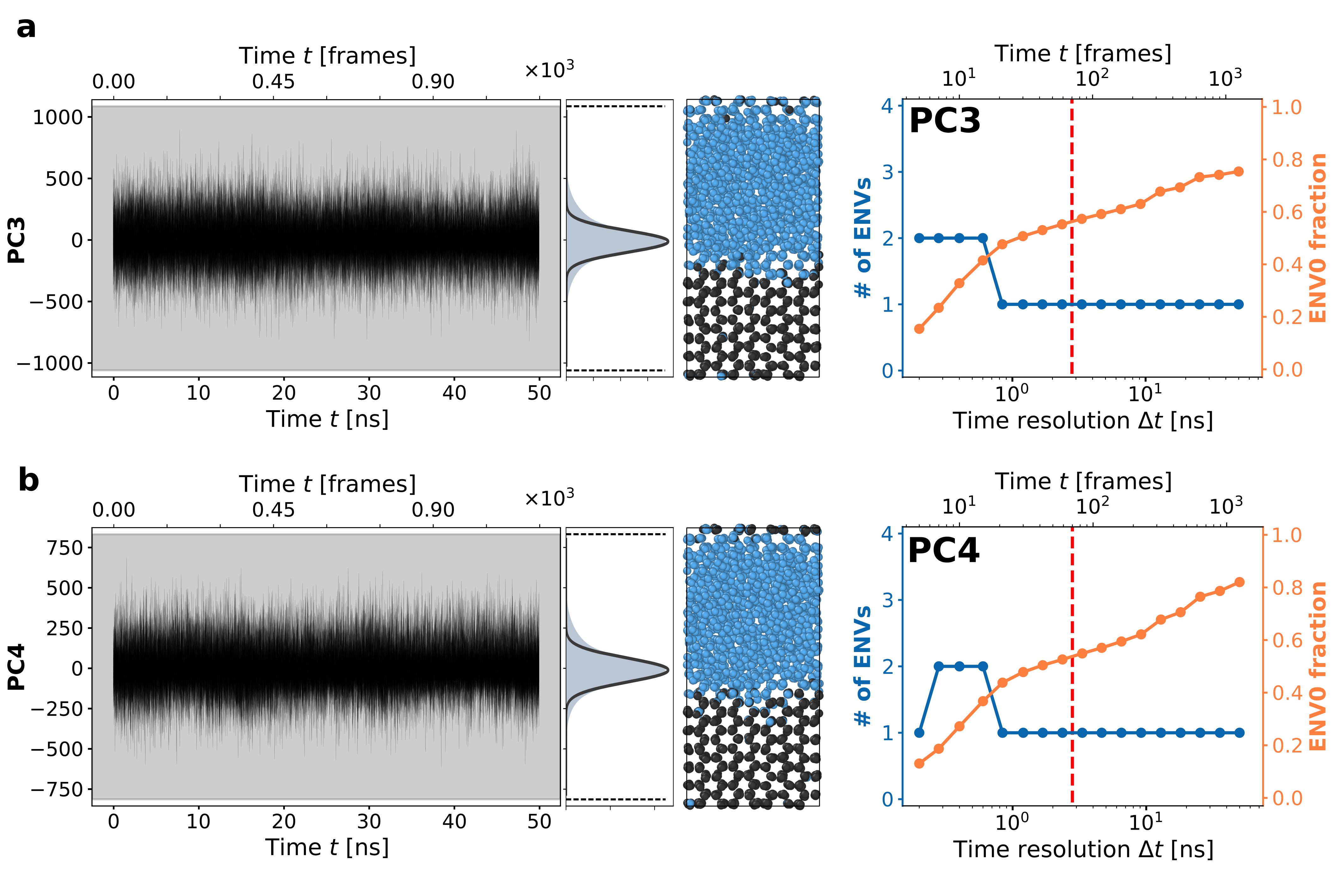}
    \caption{\textbf{Onion Clustering on PC3 and PC4 time-series data.} a) From left to right: PC3 time-series data of all molecules; kernel density estimate (KDE) of the PC3 time-series data; screenshot of the MD simulation, with molecules colored according to Onion Clustering micro-clusters; Onion output plot. b) Same as in a) but for PC4.}
    \label{figSI3}
\end{figure}

\begin{figure}[H]
\centering
    \includegraphics[width=1\textwidth]{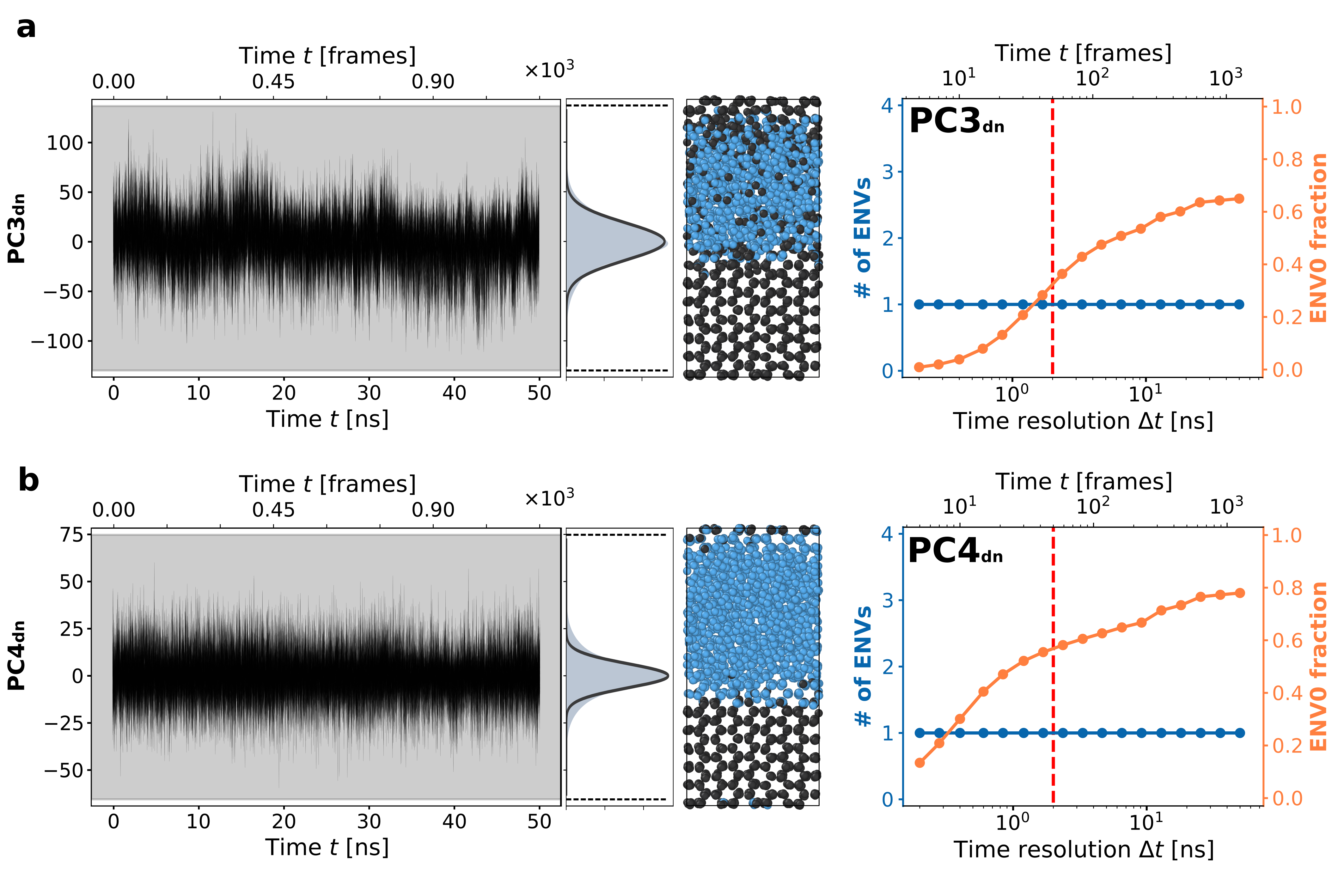}
    \caption{\textbf{Onion Clustering on denoised PC3 and PC4 time-series data.}The figure follows the same structure of Figure~\ref{figSI3}.}
    \label{figSI4}
\end{figure}

\begin{figure}[H]
\centering
    \includegraphics[width=1\textwidth]{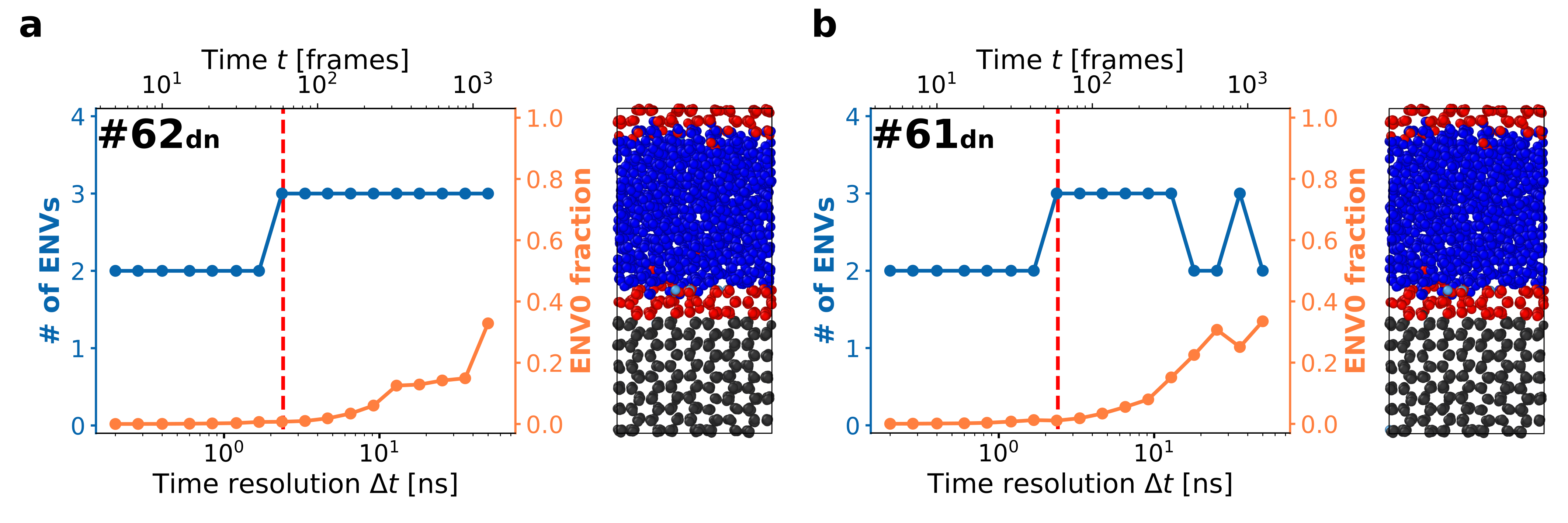}
    \caption{\textbf{Onion Clustering on denoised \(\#62\) and \(\#61\) time-series data.} a) Onion plot of \(\#62_{dn}\) and snapshots of the MD trajectory colored according to Onion micro-clusters. b) Same as a) but for \(\#61_{dn}\).}
    \label{figSI6}
\end{figure}

\begin{figure}[H]
\centering
    \includegraphics[width=1\textwidth]{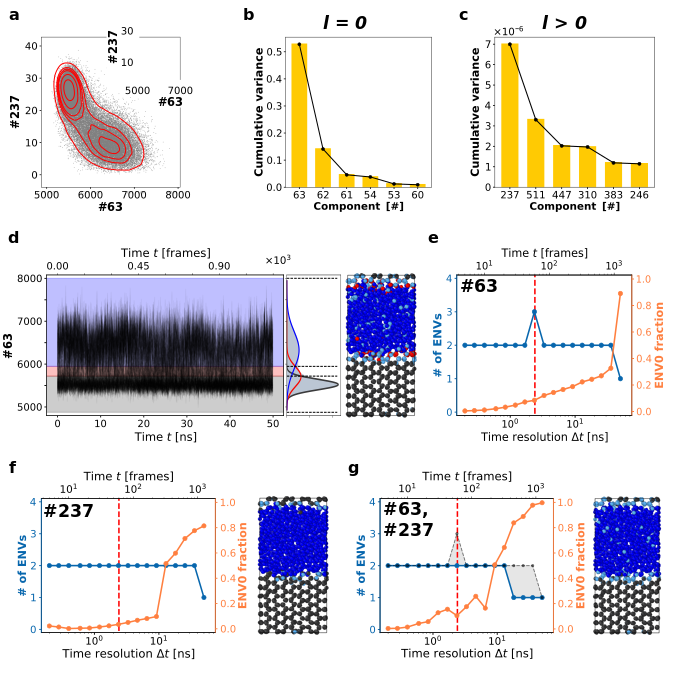}
    \caption{\textbf{Onion Clustering on raw \(\#63\) and \(\#237\) time-series data.} a) Denoised dataset projection onto components \#63 and \#237, red contour lines help visualize the data density; in the inset, static clustering distinguishes 2 environments. b) Variance of the six most significant spherical ($l=0$) SOAP components. c) Variance of the six most significant non-spherical ($l>0$) SOAP components.  d), e) Onion Clustering results of component $\#63$. f) Onion Clustering results of component $\#237$. g) Bi-dimensional Onion Clustering results of components $\#63$, $\#237$.}
    \label{figSI7}
\end{figure}

\begin{figure}[H]
\centering
    \includegraphics[width=1\textwidth]{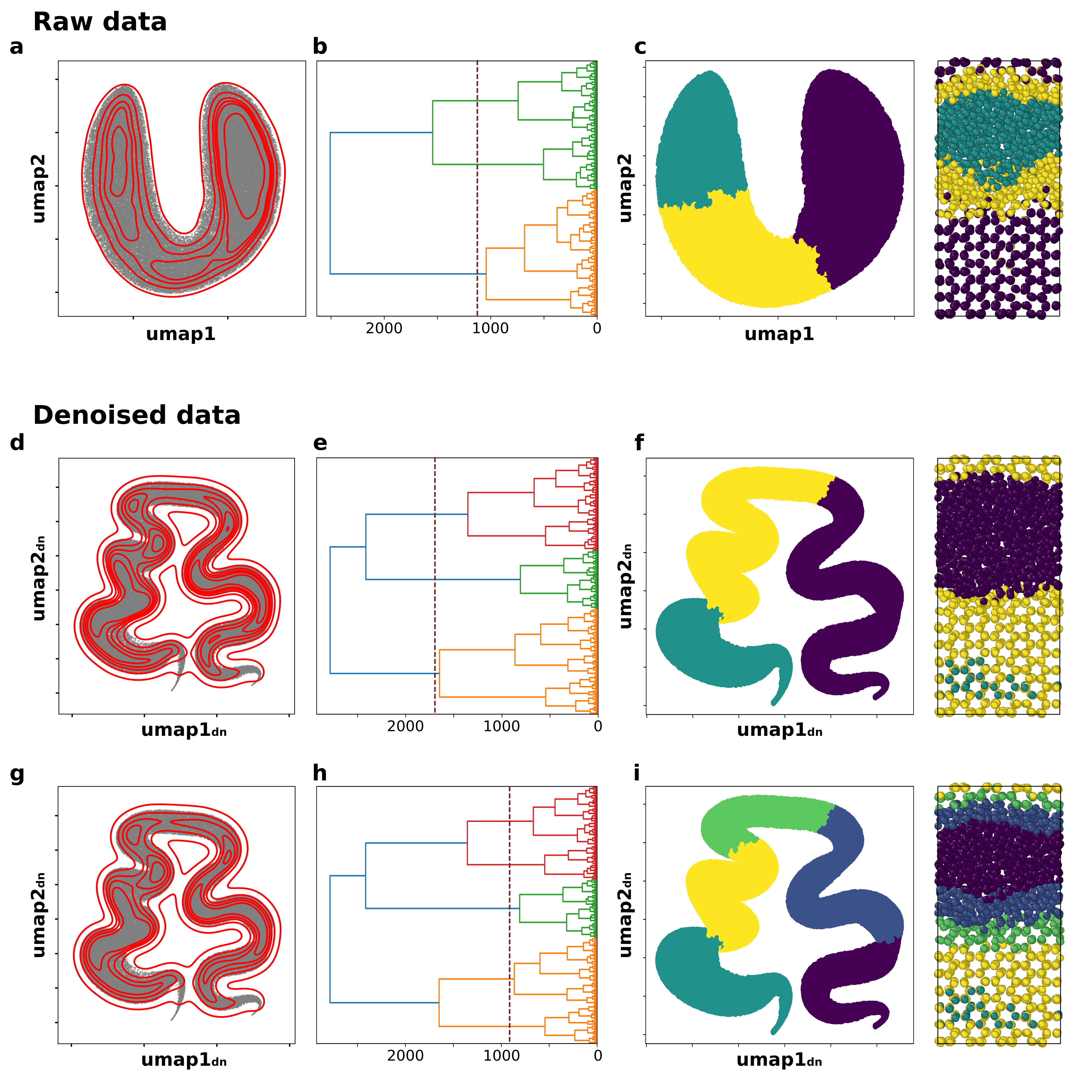}
    \caption{\textbf{UMAP results on raw and denoised SOAP dataset}. a) Projection of the raw SOAP dataset on the first two UMAP components. Red contour lines help visualize the data density. b) Dendrogram obtained by applying Hierarchical clustering on UMAP1, UMAP2. c) Right: dataset colored according to the clustering results from the Hierarchical clustering. Left: Snapshots of the MD trajectory, colored according to the micro-clusters detected. d)- i) Results of the Hierarchical clustering applied on the UMAP dataset obtained for denoised SOAP spectra. d)-f) Propose the results when cutting the dendrogram at three clusters. g)-i) Propose the results when cutting the dendrogram at five clusters.
    Time-lagged Independent Component Analysis (tICA)~\cite{molgedey1994,perez2013} performed on the same system in slightly different conditions (see Supplementary Information of Ref. \cite{Caruso2024}), combined with KMeans clustering, well distinguishes ice from water, but struggles in detecting other dynamics domains.}
    \label{figSI8}
\end{figure}

\end{document}